# Neutral Hydrogen 21cm Absorption at Redshift 0.673 towards 1504+377

C.L. Carilli

NRAO, P.O. Box O, Socorro, NM, 87801

Karl M. Menten and Mark J. Reid

SAO, 60 Garden St., Cambridge, MA, 02138

M.P. Rupen

NRAO, P.O. Box O, Socorro, NM, 87801

– 2 –

## ABSTRACT

We detect the 21 cm line of neutral hydrogen in absorption at a redshift of 0.673 towards the 1 Jy radio source 1504+377. The 1504+377 radio source is located toward the center of what appears to be an inclined disk galaxy at z = 0.674. The 21 cm absorption line shows multiple velocity components over a velocity range of about 100 km sec$^{-1}$, with a total HI column density: N(HI) = $3.8 \times 10^{19} \times (\frac{T_s}{f})$ cm$^{-2}$. The velocity-integrated optical depth of this system is the largest yet seen for redshifted HI 21 cm absorption line systems (Carilli 1995). The 21 cm absorption line is coincident in redshift with a previously detected broad molecular absorption line system (Wiklind and Combes 1996). We do not detect HI 21 cm absorption associated with the narrow molecular absorption line system at z = 0.67150, nor do we detect absorption at these redshifts by the 18 cm lines of OH, nor by the 2 cm transition of H$_2$CO.

There is no evidence for a bright optical AGN in 1504+377, suggesting significant obscuration through the disk – a hypothesis supported by the strong absorption observed. The 1504+377 system resembles the "red quasar" PKS 1413+135, which has been modeled as a optically obscured AGN with a very young radio jet in the center of a gas rich disk galaxy (Perlman et al. 1996). The presence of very bright radio jets at the centers of these two disk galaxies presents a challenge to unification schemes for extragalactic radio sources and to models for the formation of radio loud AGN.

*Subject headings:* quasars:absorption lines - radio lines: galaxies - galaxies: active, jets, ISM

## 1. Introduction

The flat spectrum radio source 1504+377 (Kühr et al. 1981) has been identified with what appears to be an inclined disk galaxy with a moderate-excitation narrow emission line spectrum at z = 0.674 (Stickel and Kühr 1994). The parent galaxy may be associated with a group of galaxies, and has a close companion galaxy at a projected distance of 4″, although the redshifts of this companion and of other possible members of the group are unknown (Stickel and Kühr 1994). The radio source shows a compact nucleus with a one-sided jet. The jet axis is oriented within $15°$ of the major axis of the galaxy, in projection (Polatidis et al. 1995, Xu et al. 1995). The radio source is variable at high frequency, showing factor two changes in total flux density at 37 GHz over four months (Wiren et al. 1992). Assuming this variation reflects the intrinsic size of the compact nucleus implies a brightness temperature of $2 \times 10^{12}$ K. The red nucleus of 1504+377 has been detected by IRAS (0.19±0.03 Jy at 60$\mu$m), and shows possible evidence for flaring in the near IR (Stickel et al. 1996).

The source 1504+377 is remarkable for three reasons. First, it is one of the rare examples of a radio loud active galactic nucleus (AGN) at the center of what appears to be a disk galaxy (Perlman et al. 1996, Wilson and Colbert 1995). Second, outside of possible flaring events, the parent galaxy of 1504+377 shows no indication of a bright AGN in the optical (Stickel and Kühr 1994, Stickel et al. 1996), suggesting 1504+377 is an example of a "red quasar", in which the active nucleus is obscured in the optical. Support for this idea comes from the recent detection of broad molecular absorption lines towards the radio source at the redshift of the parent galaxy (Wiklind and Combes 1996) and from the observation of very strong HI 21 cm absorption presented herein. And third, the presence of a bright, flat spectrum radio source at the center of an inclined, gas rich galaxy at large redshift allows for detailed absorption studies of various components of the ISM (molecular, atomic, ionized) in a galaxy at substantial look-back times.

This letter presents the detection of strong HI 21 cm absorption towards the nucleus of 1504+377, coincident in redshift with the broad molecular absorption line system and with the optical emission lines. We compare the physical conditions in the absorbing gas with those found in translucent clouds within our own galaxy, and in Centaurus A and PKS1413+135. We briefly reconsider the issues of red quasars, and radio jets in disk galaxies.

## 2. Observations

We searched for HI 21 cm absorption towards 1504+377 at the redshift of observed molecular absorption lines using the NRAO 140 ft telescope in Greenbank, WV, on the nights of May 1 to May 4, 1996.[1] The 750 to 1000 MHz feed and receiver were used. The measured system temperature was 49 K and the telescope gain at this frequency is 3.15 Jy K$^{-1}$. Two orthogonal linear polarizations were received, and position switching was employed for background and bandpass determination.

The spectrometer employed was the spectral processor (Fisher 1991), which produces a spectrum of 1024 channels for each polarization. The search involved a series of 10 hour long observations using a 10 MHz bandwidth centered at 849.20MHz. The spectral resolution was 9.7 kHz, giving a velocity resolution of 3.5 km sec$^{-1}$. On-line Doppler tracking was employed to correct topocentric motion to a heliocentric velocity frame. The gain calibration and velocity scale were checked through an observation of the redshifted HI 21 cm absorption line system towards 3C 286 (Brown and Roberts 1973). We obtained the same optical depth for the 3C 286 line as was observed previously to an accuracy of 10%, and the heliocentric corrected line frequency agreed to within 0.8 kHz, or 0.3 km sec$^{-1}$ with the published value.

---

[1] The National Radio Astronomy (NRAO) is a facility of the National Science Foundation, operated under cooperative agreement by Associated Universities, Inc.

Data for each night were calibrated, inspected for interference, and summed using the IRAF data reduction package. The summed spectra for each night were then exported to the CLASS data reduction package, where the final spectrum was generated by summing spectra from each night weighting by the integration times. The final summed spectrum had an rms noise level of 2.8 mK, close to the theoretical rms of 2.6 mK.

The total continuum flux density of 1504+377 at 849 MHz is difficult to derive using standard procedures at the 140 ft telescope due to terrestrial interference. To determine the total flux density of 1504+377 observations were made with the NRAO Very Large Array at 1.45 GHz and 0.33 GHz on May 16, 1996. The derived values were 0.98±0.02 Jy and 1.28±0.06 Jy, respectively. Interpolating between the two measurements assuming a power-law spectrum yields a flux density at 849 MHz of 1.08 Jy, implying an antenna temperature of 0.343 K at the 140 ft telescope. We use this value in the analysis below. Note that 1504+377 has not been seen to vary at frequencies below 1 GHz to a level of 10%.

## 3. Results and Analysis

The summed spectrum of redshifted HI 21 cm absorption towards 1504+377 is shown in Figure 1, together with the HCO$^+$ (2-1) absorption spectrum of Wiklind and Combes (1996). To allow direct comparison with the molecular absorption seen by Wiklind and Combes (1996), we have chosen the velocity axis in Figure 1 such that a velocity off-set of zero km sec$^{-1}$ corresponds to the heliocentric redshift of the narrow molecular absorption line system ($Z_\odot = 0.67150$).

Strong, broad absorption is seen in the HI spectrum, with the maximum optical depth of 0.32 occurring at 848.811 MHz, and a secondary maximum of 0.24 occurring at 848.897 MHz. The broad HI 21 cm absorption is coincident in redshift with the broad molecular absorption line system observed by Wiklind and Combes (1996). We do not detect absorption at the velocity of the narrow molecular absorption line system at zero velocity in Figure 1, with an implied optical depth limit ($4\sigma$) of 0.033.

Note that the peak optical depths in the HI and molecular absorption differ by 15 km sec$^{-1}$, and that shifting the spectra in velocity to line-up the absorption peaks also leads to better correspondence between the two edges of the broad absorption system. On the other hand, there is no reason to doubt either velocity scale, and differences between molecular and atomic absorption are not unusual for Galactic lines-of-sight (Lucas and Liszt 1996), hence such a shift should not be justified simply on the basis of line ratios. There is weak interference about 5 channels away from the expected redshift for the narrow absorption system in Figure 1. This interference does not affect the spectrum at zero velocity. We note however, that if one adopts the 15 km sec$^{-1}$ shift discussed above, the interference would corrupt the HI spectrum at the velocity of the narrow molecular absorption line.

The broad HI 21 cm absorption line system towards 1504+377 is well fit by three Gaussian components in velocity, as shown in Figure 1. The results for the fits are summarized in Table 1. Column 1 lists the heliocentric redshift of the line peak. Column 2 lists the central velocity, where zero velocity is defined as $Z_\odot = 0.67150$. Column 3 lists the velocity-integrated opacity for each component, and columns 4, 5, and 6 list the peak optical depth, the FWHM of the Gaussian component, and the implied HI column density, respectively. The errors include a 10% uncertainty in absolute gain determination. The velocity-integrated opacity for the total absorption is 21±2 km sec$^{-1}$, making this system the strongest of the redshifted HI 21 cm absorption line systems (in terms of velocity-integrated optical depth), as summarized in Carilli (1995). The implied total HI column density is N(HI) = $3.8 \times 10^{19} \times (\frac{T_s}{f})$ cm$^{-2}$, where T$_s$ is the spin temperature of the gas and $f$ is the HI covering factor. For the narrow molecular absorption system at Z$_\odot$ = 0.67150, the 4$\sigma$ upper limit for the HI column density at 3.4 km sec$^{-1}$ resolution is $1.9 \times 10^{17} \times (\frac{T_s}{f})$ cm$^{-2}$.

A search was made with the 140 ft telescope for absorption by the 18 cm lines of OH towards 1504+377 at the redshifted frequency of 996 MHz. No absorption was seen, to an optical depth limit (4$\sigma$) of 0.10 at 3 km sec$^{-1}$ resolution. A search was also made on July 23, 1996 with the Very Large Array for absorption by the 2 cm 2$_{11}$-2$_{12}$ transition of H$_2$CO at the redshifted frequency of 8667 MHz. No absorption was seen to an optical depth limit (4$\sigma$) of 0.005 at 3 km sec$^{-1}$ resolution. The lack of detection of OH and H$_2$CO absorption supports the conclusion of Wiklind and Combes (1996) that the low optical depth in $^{12}$CO is not due to a low covering factor, but indicates a non-saturated absorption line.

Stickel and Kühr (1995) and Stickel et al. (1996) demonstrate a sharp cut-off in the integrated spectral energy distribution (SED) for 1504+377 from the radio into the optical. Following the example of McHardy et al. (1991) for PKS 1413+135, we have fit a function to the radio through far IR SED of 1504+377 involving a power-law plus an exponential cut-off: I$_\nu \propto \nu^\alpha e^{\frac{\nu}{\nu_s}}$. We find $\alpha = -0.1$ and $\nu_s = 7 \times 10^{12}$ Hz redshifted to the rest frame of the source. For comparison, the results for 1413+135 were: $\alpha = -0.5$, and $\nu_s = 5 \times 10^{13}$ Hz.

In order to derive neutral hydrogen column densities from the absorption line data, one must determine the covering factor, f, and the spin temperature, T$_s$. Very long baseline interferometric (VLBI) images of 1504+377 at 5 GHz and 1.6 GHz reveal a compact, flat spectrum nucleus of size < 1.4 mas (= 5.4 $h^{-1}$ pc at z = 0.673); a bright, flat spectrum inner jet extending to about 10 mas (= 40 $h^{-1}$ pc); and a more diffuse component only seen at 1.6 GHz at a distance of 55 mas (= 220 $h^{-1}$ pc) from the nucleus (Polatidis et al. 1995, Xu et al. 1995), where $h \equiv \frac{H_o}{100 km sec^{-1} Mpc^{-1}}$. Using the observed spectral indices between 5

GHz and 1.6 GHz for the various components, we calculate that at 849 MHz the compact nucleus contributes about 46% to the total flux, while including the inner jet to 10 mas raises the total to 74%. We conclude that 0.46 is a likely lower limit to the covering factor, while for an HI cloud with a size corresponding to a typical Galactic giant molecular cloud of 10 pc to 50 pc (Blitz 1990), the covering factor would likely be closer to 0.74. We adopt this latter value in the analysis below. A firm lower limit to the covering factor is dictated by the peak optical depth, and by assuming an optically thick cloud, for which f $\geq$ 0.27.

Estimating the spin temperature is problematic. Detailed studies of cold atomic absorption in our galaxy, and other nearby galaxies, show spin temperature values of about 100 K (Braun and Walterbos 1992). For lack of other information we adopt this value in the analysis below.

Wiklind and Combes analyze only the velocity integrated properties of the broad absorption line system towards 1504+377. They find an excitation temperature of 16 K and a total CO column density of $6 \times 10^{16}$ cm$^{-2}$, typical values for Galactic "translucent clouds" (van Dishoeck et al. 1993). The velocity-integrated ratio of molecular to atomic columns for the broad system towards 1504+377 is: $\frac{N(CO)}{N(HI)} = 1.2 \times 10^{-5}$. For comparison, the redshifted absorption system towards PKS 1413+135 has a total HI column density of $1.3 \times 10^{21}$ cm$^{-2}$, assuming $T_s = 100$ K and $f = 1$, and an integrated CO to HI column density ratio of $8 \times 10^{-6}$ assuming a CO excitation temperature of 10 K (Carilli et al. 1992, Wiklind and Combes 1994). The associated molecular and atomic absorbing system seen towards the nucleus of the nearby "dust-lane" radio galaxy Centaurus A shows: N(HI) = $5 \times 10^{21}$ cm$^{-2}$, and $\frac{N(CO)}{N(HI)} = 2 \times 10^{-5}$ (Israel et al. 1990, Israel et al. 1991, Seaquist and Bell 1990, Eckart et al. 1990, van der Hulst, Golish, and Haschick 1983).

As in the case of PKS 1413+135 (Wiklind and Combes 1994, Carilli et al. 1992), the velocity width of the HI absorbing gas totally encompasses the molecular absorption in 1504+377. This may indicate a longer pathlength through the HI gas than through the molecular gas, and/or that the background source has a larger spatial extent at low frequency, hence providing a larger 'pencil-beam.'

## 4. Discussion

Wiklind and Combes (1996) adopt z = 0.67150 as the systemic velocity of the parent galaxy, assuming that the narrow system is a molecular cloud in the outer disk of the galaxy for which the predominant motion will be transverse to our line-of-sight, while the broad system is a cloud complex associated with the AGN. However, the optical redshift of the galaxy is well determined from six narrow emission lines, giving a redshift of 0.674±0.001, where the error bar indicates the full range of measured redshifts for the different lines (Stickel and Kühr 1994). Within this error, the optical lines have the same redshift as the HI 21 cm absorbing gas and the broad molecular absorbing system. We feel it likely that this higher redshift is the systemic redshift of the parent galaxy, and that the narrow molecular absorption line is blue-shifted relative to the optical galaxy by 330 km sec$^{-1}$.

The velocity spread for the broad HI and molecular absorption line system towards 1504+377 is much larger than expected for a single giant molecular cloud, and is comparable to the width of the associated absorbing system seen towards the nucleus of Centaurus A. Israel et al. (1990, 1991) model the absorbing material towards Centaurus A as a circum-nuclear molecular disk (size ≈ 100 pc) in which the high velocities may be explained by elliptical orbits for the clumpy torus material. Imaging the absorbing HI against the extended radio continuum source with milli-arcsec resolution could test such a model for 1504+377.

Lucas and Liszt (1996) and Liszt and Lucas (1996) derive relationships between the atomic and molecular absorption lines for a sample of diffuse or translucent Galactic molecular clouds. The ratio of peak optical depth in HCO$^+$ versus HI, $R \equiv \frac{\tau(HCO^+)}{\tau(HI)}$, varies as $0.1 \leq R \leq 6$, for their 13 molecular clouds. They explain such variations with a 'layered cloud' model, where the outer parts of the cloud are mostly atomic and the molecular fraction increases with decreasing radius. For the broad line system in 1504+377 we find $R \approx 1$ at the two peaks in the HI profile, and the variation of this ratio across the entire profile is within the range observed for Galactic clouds by Lucas and Liszt.[2]

The $4\sigma$ limit to HI absorption by the narrow line system towards 1504+377 implies $R > 9$. The HI paucity in this narrow molecular line system towards 1504+377 resembles that seen in the low latitude Galactic absorbing cloud towards 0727-115 at $v_{LSR} = 35$ km sec$^{-1}$ (Lequeux, Allen, and Guilloteau 1993, Dickey et al. 1983), which has a $4\sigma$ lower limit of $R > 7$ (Dickey et al. 1983, Lucas and Liszt 1986). The lack of associated HI absorption in these two systems could result either from warmer HI than is typical for diffuse clouds, or from a lower degree of molecular dissociation, perhaps indicating a weaker interstellar UV radiation field (Lequeux et al. 1993).

The large out-flow velocity of 330 km sec$^{-1}$ of the narrow molecular absorbing cloud relative to systemic is inconsistent with a quiescent cloud in the outer disk of the parent galaxy, and is even outside the velocity range observed for associated HI absorption towards the nuclei of 43 nearby Seyfert galaxies (Dickey 1986). The existence of a small galaxy just 4″ (= 16 $h^{-1}$ kpc) from the parent galaxy of 1504+377 raises the possibility of high velocity gas due to tidal interaction. High resolution optical imaging is required to determine whether the two galaxies are interacting gravitationally.

---

[2]Note that Lucas and Liszt observed the 1-0 transition of HCO$^+$ while in 1504+377 the 2-1 transition was observed. For HCO$^+$ excitation at the microwave background temperature of 4.57 K at z = 0.673 the expected optical depths in the two transitions will be the same to within 10%.

The 1504+377 system is similar to the well studied red quasar PKS 1413+135 at z = 0.25 (McHardy et al. 1991, McHardy et al. 1994, Perlman et al. 1996, Stocke et al. 1992, Carilli, Perlman, and Stocke 1991, Wiklind and Combes 1994). Both are radio loud AGNs with pc-scale jets at the centers of what appear to be inclined disk galaxies. Both show strong HI 21 cm and molecular absorption toward the radio source. Both show substantial high frequency radio variability indicating high brightness temperatures. And both show a cut-off in the radio-to-optical spectrum, indicating substantial obscuration toward the AGN. There are a few significant differences between the two sources: the parent galaxy of 1504+377 may be in a group, while 1413+135 is not; the 1504+377 galaxy shows narrow optical emission lines, while 1413+135 does not; and the 1504+377 radio source has a one-sided jet, while 1413+135 has a two-sided jet.

For 1413+135 and 1504+377 the detection of a substantial molecular and neutral atomic ISM argues for a parent galaxy type later than S0. For 1413+135 a recent HST image reveals a prominent dust lane, and McHardy et al. (1994) were able to show quite convincingly that the galaxy type was later than S0, through surface brightness modeling.

Perlman et al. (1996) model 1413+135 as a very young radio jet (age $\approx 10^4$ yr) in the center of a gas-rich disk galaxy. From the lack of optical emission lines and the lack of a bright IR nucleus they conclude that either the radio source is highly relativistically beamed, thereby lowering the total intrinsic luminosity, or that the obscuration is highly anisotropic, i.e. that most of the AGN radiation escapes out of the plane of the galaxy.

The Perlman et al. model of a young, optically obscured, radio-loud AGN at the center of a gas-rich disk galaxy applies very well to 1504+377. Indeed, the presence of emission lines in 1504+377 alleviates one of the difficulties encountered in 1413+135, while the one-sided jet in 1504+377 lends itself more easily to beaming models. Perlman et al. discuss the possibility that the PKS 1413+135 radio source may be a background source, possibly gravitationally lensed by the galaxy at z = 0.25. While they cannot rule out this possibility, they show that the radio morphology of the system – in particular the lack of multiple imaging on scales from 100 mas to a few arc seconds, is inconsistent with lensing (see also McHardy et al. 1994). A similar argument applies in the case 1504+377, with the additional fact that the jet in 1504+377 is significantly less distorted than that seen in 1413+135.

The most significant difficulty with this model for both 1504+377 and 1413+135 is the presence of a radio-loud AGN in a disk galaxy. Low level radio emission is seen from some nearby Seyferts, however the typical radio luminosities of AGN in disk galaxies are about four orders magnitude below that seen from 1413+135 and 1504-377. The radio luminosities of these two sources are comparable to those expected for radio loud AGN in elliptical galaxies (Wilson and Colbert 1995). A possible solution to this dilemma is relativistic beaming of the radio emission. However, to recover four orders of magnitude in apparent luminosity requires a jet oriented within $4^o$ degrees of our line-of-sight, using a typical jet Lorentz factor of 4 (Vermeulen 1996). For 1504+377 this would imply a de-projected jet length larger than 3 $h^{-1}$ kpc, with the jet oriented close to the plane of the galaxy. Such a long jet propagation distance through the plane of the galaxy without disruption or change of direction appears unlikely (Colbert et al. 1996). Also, given that beaming requires a relativistic jet, the fundamental problem remains of having an AGN with a high mechanical luminosity in a disk galaxy, thereby violating some current models of AGN formation (Wilson and Colbert 1995, Urry and Padovani 1995).

We would like to thank M. McKinnon, R. Fisher, R. Maddelena, and the staff at the 140' for their help with these observations, and H. Liszt, E. Perlman, and J. Stocke for useful discussion. This research made use of the NASA/IPAC Extragalactic Data Base (NED) which is operated by the Jet propulsion Lab, Caltech, under contract with NASA.

## References

Blitz, L. 1990, in *The Evolution of the Interstellar Medium*, ed. L. Blitz (ASP: San Francisco), 273

Brown, R.L. and Roberts, M. 1973, Ap.J. (letters), 184, L7

Braun, R. and Walterbos, R. 1992, 386, 120

Carilli, C.L. 1995, J. Astrophys.& Astr. (Supplement), 16, 163

Carilli, C.L., Perlman, E.S., and Stocke, J.T. 1992, Ap.J. (Letters), 400, L13.

Colbert, E.J., Baum, S., Gallimore, J., O'Dea, C., and Christensen, J. 1996, Ap.J., 467, 551

Combes, F. and Wiklind, T. 1996, in *Cold Gas at High Redshift*, eds. M. Bremer, P. van der Werf, H. Rottgering, and C. Carilli (Kluwer: Dordrecht).

Dickey, J.M. 1986, Ap.J., 300, 190.

Dickey, J.M., Kulkarni, S.R., van Gorkom, J.H., and Heiles, C. 1983, Ap.J. (Supplement), 53, 623

van Dishoeck, E.F., Black, J., Draine, B., and Lunine, J. 1993, in *Protostars and Planets III*, ed. E. Levy and J. Lunine, p. 163

Eckart, A. et al. 1990, Ap.J. 365, 522

Fisher, Rick 1991, *Spectral Processor Manual*, NRAO, Green Bank, WV.

van der Hulst, J.M., Golish, W., and Haschick, A. 1983, Ap.J. (letters), 264, L37

Israel, F.P., van Dishoeck, E.F., Baas, F., de Graauw. T., and Phillips, T. 1991, A&A (letters), 245, L13

Israel, F.P., van Dishoeck, E.F., Baas, F., Koornneef, J., Black, J., and de Graauw. T. 1990, A&A, 227, 342

Kühr, H., Witzel, A., Pauliny-Toth, I.I., and Nauber, U. 1981, A&A (Supplement), 45, 367

Lequeux, J., Allen, R.J., and Guilloteau, S. 1993, A&A (letters), 280, L23.

Liszt, H. and Lucas, R. 1996, A&A, in press.

Lucas, R. and Liszt, H. 1996, A&A, 307, 237.

McHardy, I.M., Abraham, R.G., Crawford, C.S., Ulrich, M.-H., Mock, P.C., and Vanderspeck, R.K. 1991, M.N.R.A.S., 249, 742.

McHardy, Ian M., Merrifield, M.R., Abraham, R.G., and Crawford, C.S. 1994, M.N.R.A.S., 268, 681

Perlman, E.S., Carilli, C.L., Stocke, J.T., and Conway, John 1996, A.J., 111, 1839.

Polatidis, A. et al. 1995, Ap.J. (supplement), 98, 1

Seaquist, E.R. and Bell, M.B. 1990, Ap.J., 364, 94

Spitzer, Lyman Jr. 1978, *Physical Processes in the Interstellar Medium*, (Wiley: New York)

Stickel, M. and Kühr, H. 1994, A&A (Supplement), 105, 67.

Stickel, M., Rieke, G.H., Kühr, H., and Rieke, M.J., 1996, Ap.J., 468, 556.

Stocke, J.T., Wurtz, R., Wang, Q., Elston, R., Jannuzi, B.T., and Deiker, S. 1992, Ap.J. (letters), 400, L17.

Urry, C.M. and Padovani, P. 1995, P.A.S.P. 107, 803

Vermeulen, R. 1996, Pub. N.A.S., 92, No. 5, 11385

Wilson, A.S. and Colbert, E.J. 1995, Ap.J., 438, 62

Wiklind, Tommy and Combes, Francoise 1996, A&A, in press

Wiklind, T. and Combes, F. 1994, A&A (letters), 286, L9

Wiren, S., Valtaoja, E., Terasranta, H., and Kotilainen, J. 1992, A.J., 104, 1009

Xu, W., Readhead, A.C.S., Pearson, T.J., Polatidis, A., and Wilkinson, P.N. 1995, Ap.J. (supplement), 99, 297

Table 1. Gaussian Model Parameters

| $Z_\odot$ | Velocity km sec$^{-1}$ | Integrated Opacity km sec$^{-1}$ | Optical Depth | FWHM km sec$^{-1}$ | Column Density $\times \frac{T_s}{f}$ 10$^{19}$ cm$^{-2}$ |
|---|---|---|---|---|---|
| 0.67329±0.00002 | 322±5 | 6.6±0.9 | 0.048±0.006 | 130±14 | 1.2 |
| 0.67324±0.00001 | 312±2 | 2.9±0.3 | 0.16±0.017 | 17.1±0.9 | 0.5 |
| 0.67343±0.00001 | 347±2 | 11.5±1.2 | 0.27±0.027 | 39.4±1.4 | 2.1 |

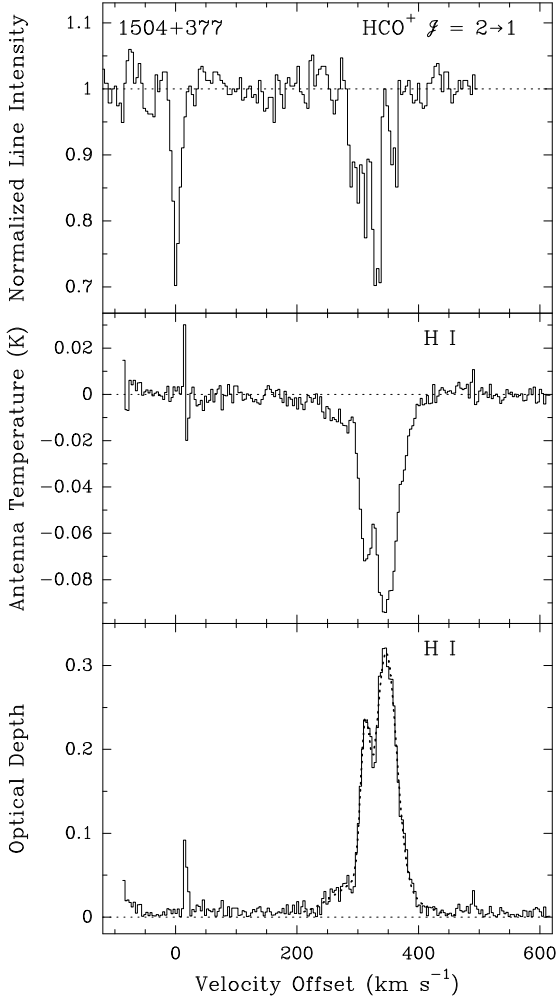

Fig. 1.— The upper frame shows the redshifted HCO$^+$ (2-1) absorption spectrum of 1504+377 (Wiklind and Combes 1996). The HCO$^+$ spectrum has been normalized such that the source flux density corresponds to unit intensity. The middle frame shows the redshifted HI 21 cm absorption spectrum towards 1504+377. The bottom frame shows the HI optical depth spectrum plus the 3 component Gaussian model fit (dotted line). The parameters for the fit are given in Table 1. Following the convention of Wiklind and Combes (1996), the zero point on the velocity scale corresponds to the heliocentric redshift of the narrow molecular absorption line system ($Z_\odot = 0.67150$). Note that the HI spectrum is corrupted by terrestrial interference at a velocity of about +15 km sec$^{-1}$.